\documentclass{mn2e}
\usepackage{color}
\usepackage{graphicx}
\usepackage{dcolumn}
\usepackage{bm}
\usepackage{lscape}

\title[Comment on Kormendy et al.\ (2011)] 
{Comment on Kormendy et al.\ (2011, Nature, 469, 374)}

\author[Graham et al.]
{Alister W.\ Graham$^1$\thanks{AGraham@astro.swin.edu.au}
\\
$^1$ Centre for Astrophysics and Supercomputing, Swinburne University
of Technology, Hawthorn, Victoria 3122, Australia.\\
}

\begin{document}
\label{firstpage}
\maketitle

\begin{abstract}

Comments on Kormendy, Bender \& Cornell (2011, Nature, 469, 374;
arXiv:1101.3781) are presented. 
A broader historical perspective is provided before a number of 
scientific concerns regarding their data and conclusions are 
discussed. 

\end{abstract}


\section{Historical Background}

The non-compliant nature of certain disc galaxies in the (black hole
mass)-(bulge velocity dispersion) diagram, known as the
$M_{\mathrm{bh}\,}$-$\,\sigma$ diagram (Figure 2 in Kormendy et al.'s Nature
article), was first publicly announced three years ago, in January 2008 at the
American Astronomical Society Meeting in Austin Texas.  The associated journal
paper (Graham 2008a), connected with the conference presentation (Graham
2007), revealed how barred galaxies can be offset from the
$M_{\mathrm{bh}\,}$-$\,\sigma$ relation defined by non-barred and elliptical
galaxies.  This discovery was simultaneously framed in terms of offset
pseudobulges by Hu (2008), with Hu's pseudobulge sample effectively a barred
galaxy sample\footnote{Only one of Hu's (2008) nine ``pseudobulge'' galaxies 
 was not barred.}.  Graham (2008a) reported that there was a 0.1 to 0.01 per cent 
probability of the galaxy sample being offset by chance.  The deviant nature
of these barred/pseudobulge galaxies, which has included the Milky Way, 
was again the focus in Graham (2008b),
Graham \& Li (2009), Gadotti \& Kauffmann (2009), Greene et al.\ (2010), and
discussed extensively in mid 2010 by Graham et al.\ (2010).  This latter work
expanded the sample of galaxies used in 2008-2009 from $\sim$50 to 64 (including 39
disc galaxies, 20 of which are barred galaxies) and constructed new relations
after having identified and corrected an additional sample bias --- an
artificial black hole mass floor in the data set --- affecting past studies.

The classical $M_{\mathrm{bh}\,}$-$\,\sigma$ relation was constructed using
galaxies of all morphological type (Ferrarese \& Merritt 2000; Gebhardt et
al.\ 2000; Merritt \& Ferrarese 2001; Tremaine et al.\ 2002).  To distinguish
from this, Graham (2008a) introduced the barless $M_{\mathrm{bh}\,}$-$\,\sigma$
relation and the elliptical-galaxy $M_{\mathrm{bh}\,}$-$\,\sigma$ relation (see
also Hu 2008 in regard to the latter), 
and reported that "Removal of the seven barred galaxies from the Tremaine et al.\
(2002) set of 31 galaxies gives a barless $M_{\mathrm{bh}\,}$-$\,\sigma$ relation with an intrinsic
scatter of 0.17 dex (cf.\ 0.27 dex for the 31 galaxies) and a total scatter of
0.25 dex (cf.\ 0.34 dex for the 31 galaxies)".

The $M_{\mathrm{bh}\,}$-$\,\sigma$ relation was predicted (Silk \& Rees 1998;
Haehnelt, Natarajan \& Rees 1998; Fabian 1999, 2010) before it was observed. 
The Silk \& Rees (1998)
$M_{\mathrm{bh}\,}$-$\,\sigma$ relation with a slope of 5 is now observationally
supported by both the barless $M_{\mathrm{bh}\,}$-$\,\sigma$ relation and the
elliptical-only $M_{\mathrm{bh}\,}$-$\,\sigma$ relation (Graham et al.\ 2010;
see also Ferrarese \& Ford's 2005 classical $M_{\mathrm{bh}\,}$-$\,\sigma$ relation).

As revealed by Novak, Faber \& Dekel (2006), it is not yet established which
physical property of the host bulge best correlates with the mass of the
central black hole.  What is known is that the scatter about the classical
$M_{\mathrm{bh}\,}$-$\,\sigma$ relation has increased as the number of barred,
and likely pseudobulge, galaxies has increased (Hu 2008; Graham 2008b;
G\"ultekin et al.\ 2009; Graham et al.\ 2010).  Graham (2008a) wrote "Bar
instabilities are believed to lead to the formation of pseudobulges. Such
evolution may have resulted in (pseudo)bulges with an increased velocity
dispersion and luminosity but a relatively anemic SMBH (unless it also grew
during the formation of the pseudobulge). If the barred galaxies do indeed
have discrepantly low SMBH masses rather than high $\sigma$-values, they
should also appear as systematic outliers in the $M_{\mathrm{bh}\,}$-$L$
diagram."  The work by Graham and collaborators does not rule out this
possibility which is what the Nature paper in question has attempted to
answer.

\section{Cautionary remarks on pseudobulge identification and luminosity}

Pseudobulges are notoriously hard to identify, and there is not yet a
consensus as to how to define them.  For example, at odds with the Nature
article, Peebles' (2011) review of the article reports that a pseudobulge "is
a concentration of starlight near the centre of [a] galaxy, but in the disk,
not extending above it as do stars in a bulge".  

While the detection of 
rotating bulges, pseudo or not, goes back a long time (e.g.\ Babcock 1939;
Rubin, Ford \& Kumar 1973; Pellet 1976; Bertola \& Capaccioli 1977; Peterson
1978; Mebold et al.\ 1979; Rubin 1980), some classical bulges, just like
low-luminosity elliptical galaxies, are expected to have significant rotation
(Naab, Khochfar \& Burkert 2006; Bekki 2010).  Classical bulges can also
appear to rotate due to the presence of a bar (e.g.\ Babusiaux et al.\ 2010).
Furthermore, classical bulges can have S\'ersic (1968) indices (a measure of
how centrally concentrated their stellar light is, see Graham \& Driver 2005
and references therein) less than 2, just as low-luminosity elliptical
galaxies do (e.g.\ Caon et al.\ 1993; Young \& Currie 1994; Scannapieco et
al.\ 2010; Graham 2010 and references therein).  

To further complicate matters, classical and pseudobulges can exist within the
same galaxy, such as the case of NGC 2787 (Erwin et al.\ 2003) used in the
Nature article.  The classical bulge in this galaxy has a
magnitude roughly 1 mag fainter than the pseudobulge, yet no distinction is
made in the Nature article's Figure 1 which would have revealed, at odds with
the Nature article's premise, that the classical bulge, rather than the
pseudobulge, is the outlying point.\footnote{Curiously, it is of interest to
note that the alleged pseudobulges in Greene, Ho \& Barth (2008, their
figure~7) are consistent with the $M_{\rm bh}$-$M_{\rm bulge}$ relation from
H\"aring \& Rix (2004), while their classical bulge data is inconsistent
with this relation.}

Figure 1 in the Nature article reveals that only 3 or 4 alleged pseudobulges appear to depart from
the $M_{\mathrm{bh}\,}$-$L$ relation defined by the other galaxies, while the
remaining 7 or 8 alleged pseudobulges, i.e. the bulk of them, agree with the
main relation.  This would appear to support the notion of the coevolution of black holes and
pseudobulges, at odds with the article's conclusion.  
This observation might instead reflect that the actual number of pseudobulges
has been significantly over-estimated due to the use of questionable selection
criteria and/or reveal a problem with the assigned bulge luminosities.  One of
the few offset galaxies is NGC 1068, an SB galaxy with a large 3 kpc bar
oriented at a position angle of approximately 45 degrees (Scoville et
al.\ 1988; Thronson et al.\ 1989).  Given the connection between pseudobulges
and the presence of large-scale bars (and nuclear bars), additionally
modelling the bar light, along with the S\'ersic-bulge plus the exponential-disc
light, would be appropriate given that the galaxy samples now contain many
barred galaxies.  This, however, has not been done.  

Figure 1 in the Nature article has used a bulge-to-total
ratio of 0.11 for the barred (pseudobulge) galaxy NGC~3227, while not assigning the bar
light to the bulge is known to give a ratio of 0.068 (Gadotti 2008), i.e.\ 
the bulge is actually half a magnitude fainter than assumed.
Similarly, the barred galaxy NGC~4596 has been assigned a ratio of
0.3 in the article, where as Laurikainen et al.\ (2005) revealed that the
bulge-to-total ratio is 0.13 when 
the bar light is excluded, this amounts to a difference of nearly 1 mag in
Figure 1 of the Nature article, and explains much of the apparent offset 
nature of these galaxies in the $M_{\mathrm{bh}\,}$-$L_{\rm bulge}$ diagram. 

A proper decomposition of the light is important to 
prevent over-estimating the bulge flux from not only large-scale bar light
that is assigned to the bulge, but 
from unmodelled nuclear bars and star clusters which can increase the apparent
S\'ersic index of the bulge and consequently its derived luminosity (e.g.\
Balcells et al.\ 2003).
Moreover, given the coexistence of massive black holes in dense nuclear star
clusters at the low mass end of the $M_{\mathrm{bh}\,}$-$\sigma$ diagram 
(Seth 2008, Gonzalez-Delgado 2008; Graham \& Spitler 2009 and
references therein), one is not looking at the full picture if one ignores 
these compact nuclei with masses up to $10^7 M_{\odot}$.  Accounting for their
mass may well partially explain the apparent offset nature of some (pseudobulge)
galaxies at the low-mass end of the $M_{\mathrm{bh}\,}$-$\sigma$ diagram.

Traditionally, all bulges were assumed to have the same concentration; the
same S\'ersic index with a value of 4 was used to describe their radial flux
distribution.  We now know this was wrong (e.g.\ Andredakis 1994; de Jong 1996;
Balcells et al.\ 2003).  Unfortunately the literature remains full of
bulge-to-total ($B/T$) flux ratios which are too high.  The average ratio for
(classical bulges abundant among) the S0 galaxy population is $\sim$0.25
(e.g.\ Balcells et al.\ 2003; Laurikainen et al.\ 2005, 2007, 2010; Graham \& Worley
2008). It is therefore somewhat concerning to find half of the classical
bulges in Table 1 of the article with $B/T > 0.5$.  At the other extreme is the unaccounted for
excess central flux seen in the late-type spiral galaxy NGC 4395 (see Graham \&
Spitler 2009).  That is, at least one of the two allegedly pure-disc
galaxies with a black hole mass that is not consistent with a value of zero
appears to have a bulge, pseudo or otherwise, of the size expected for its
morphological type.

Only with reliable bulge luminosities will we know if pseudobulge and barred
galaxies follow the same (black hole mass)-(bulge luminosity) relation as
elliptical galaxies and non-barred galaxies.  Importantly, if no clear offset
of the former relative to the distribution of the latter in this diagram is
found, it would suggest that these disc galaxies, or at least the majority of
them, have bulges which coevolve with their black holes.  The already
established offset nature of (some of the) galaxies with bars and pseudobulges
in the $M_{\mathrm{bh}\,}$-$\,\sigma$ diagram would then suggest an issue with
the measured dynamics, i.e. the stellar "velocity dispersions" of the host
bulges.  This may arise due to differing levels of interference from the
dynamics of the bars whose light signal is mixed with that of the bulge (an
issue explore by Graham et al.\ 2010).
Furthermore, if bar dynamics deviate significantly from the often assumed
axisymmetric, rather than triaxial, stellar orbits used to constrain the black
hole mass, then this could also contribute to potential offsets of barred
galaxies in both the $M_{\mathrm{bh}\,}$-$\,\sigma$ and
$M_{\mathrm{bh}\,}$-$\,L$ diagram.

\end{document}